\documentclass[aps,prl,showpacs,preprint]{revtex4-1}

\usepackage{amssymb}

\begin{document}

\title{A quantum loophole to Bell nonlocality} 

\author{V\'{\i}ctor Romero-Roch\'in} 

\email{romero@fisica.unam.mx}

\affiliation{Instituto de F\'{\i}sica, Universidad
Nacional Aut\'onoma de M\'exico. \\ Apartado Postal 20-364, 01000 M\'exico D.
F. Mexico.}

\date{\today}

\begin{abstract}
We argue that the conclusion of Bell theorem, namely, that there must be spatial non-local correlations in certain experimental situations, does not apply to typical individual measurements performed on entangled EPR pairs. Our claim is based on three points, (i) on the notion of quantum {\it complete measurements}; (ii) on Bell results on local yet distant measurements; and (iii) on the fact that perfect simultaneity is banned by the quantum mechanics. We show that quantum mechanics indicates that, while the measurements of the pair members are indeed space-like separated, the pair measurement is actually a sequence of two complete measurements, the first one terminating the entanglement and, therefore, the second one becoming unrelated to the initial preparation of the entangled pair. The outstanding feature of these measurements is that  neither of them violates the principle of locality. We discuss that the present measurement viewpoint appears to run contrary to the usual interpretation of ``superposition" of states with its concomitant ``collapse" on measurement. 
\end{abstract}

\maketitle

{\it Introduction.} Perhaps the most transcendental repercussion of Bell theorem \cite{Bell1,Bell2} is the implication that Nature, and quantum mechanics as its best describing theory, are both spatially non-local. Bell theorem appears to assert that any theory that obeys relativistic causality and that assumes spatial locality, imposes limits on the correlations of measurements performed at space-like separations. These limitations can be mathematically expressed in the form of Bell inequalities \cite{Bell1,Ineq}. As it was originally pointed out by Bell \cite{Bell1}, those inequalities are not satisfied by many quantum entangled states and, therefore, the conclusion is that the World and quantum mechanics are non-local in general. Initiating with the seminal experiments by Aspect et al. \cite{AspectEPR}, there has been a plethora of experiments \cite{expEPR} that confirm that the predictions of quantum mechanics hold true, although there may still be so-called experimental loopholes that would bound such an assertion. Despite these theoretically and experimentally widely supported contentions that have verified the violations of Bell inequalities and, hence, the purportedly non-locality of Nature, the goal of this article is to argue that quantum mechanics by itself provides a loophole to Bell theorem. That is, we shall maintain that while Bell theorem is true, it does not necessarily apply to individual typical measurement of an entangled Einstein-Podolsky-Rosen (EPR) pair \cite{EPR,Bohm}. \\

In order to present our ideas, we will first indicate the meaning we are given to the statement of ``complete measurement". Although we believe this is standard knowledge in the theory of quantum mechanics, it is certainly not emphasised in regular textbooks. As we will discuss, this clarification of completeness of a measurement is synonymous to giving a meaning or interpretation to a pure quantum state of a closed system. We will show that while there is a precise and definite number of noncommuting variables that specify a quantum state, the number of the needed measured variables to achieve certainty on all of them  may be {\it smaller}. Since such a minimum number of measurements depends on a variety of situations, we will emphasise that a general prescription to determine such a number cannot be specified. It must be specified case by case. In this light, we shall find that one salient characteristic of an entangled state is that a measurement of a single member of the pair constitutes a complete measurement of the whole pair. While this may be ``bothersome", and it seems it was for EPR, because we can find the value of a quantum variable without interfering with it, in apparent contradiction with the tenants of quantum mechanics, it is actually an inevitable consequence of the Uncertainty Principle, as Bohr explained in his reply \cite{Bohr} to EPR. This process, however, using Bell arguments, can be shown to be susceptible of being explained in local terms. \\

Regarding an actual experiment in which physical devices are set to measure both members of an entangled pair, and since a complete measurement signifies having reached full certainty of a state of the quantum system under investigation, we argue that one member of the pair is {\it always} measured before the other one. Hence, this single one measurement yields full certainty of the states of both members of the pair. At this stage, the predictions of the entangled state ceases for the pair and, at the same time, it constitutes a preparation of a  known initial state for the member of the pair not yet registered by the other measuring device. This second member of the pair thus faces a later measurement whose results are no longer correlated with the initial preparation of the entangled state.  The ensuing prediction of the expectation value of many similar measurements of the pair is the same as the usual one, but its elucidation is very different from the typical interpretation. By understanding the measurement process in this way, one immediately sees that Bell proposal of the correlation function of the two measurements no longer applies. Bell correlation can be applied to the {\it first} complete measurement only, but, as stated above, such a case does not violate locality. An apparent difficulty with this explanation arises if we claim that a perfect {\it simultaneous} measurement of the pair can be performed. We contest that perfect simultaneity requires the knowledge of the {\it velocity} of a reference frame or of a particle, in either case requiring a concept foreign to quantum mechanics. At best, simultaneity can only be specified within the limits posed by the uncertainty of position and/or time measurements. On the other hand, it cannot be asserted that in real experiments the measurements of the many pairs analysed are all simultaneous pair by pair. Indeed, the coincidence counts in actual experiments is not for assuring perfect simultaneity but for ensuring that the measurements correspond to entangled pairs. As a matter of fact, even if it is a demanding experiment, we are all sure that if the measurements were made at well separated different times, in the laboratory reference frame, the experimental correlations would be the same. This would simply be a corroboration that the measurements are space-like separated.\\

{\it On complete measurements.} In many discussions of quantum mechanics there is a tendency of being highly ``abstract". That is, one speaks of operators $\hat {\cal O}$ and state vectors $| \psi \rangle$ in a Hilbert space ${\cal H}$, followed by the mathematical rules of linear algebra, and their interpretation in terms of ``superposition" of states and their ``collapse" after the measurement of an observable represented by a Hermitian operator $\hat A$. Time evolution of the state $| \psi \rangle$ is incorporated by a unitary transformation using an also abstract Hamiltonian operator $\hat H$. What is missing in these, otherwise correct discussions, is the emphasis 
that for any resemblance with the real world, one needs to specify the type and number of degrees of freedom that compose the system under study, as well as their interactions among themselves and with any possible external field. It is peculiar to note that in usual discussions of classical mechanics the specification of the system degrees of freedom is always at the forefront, but certainly it is not the rule in addressing quantum mechanics as a general theory. The stipulation of the degrees of freedom is crucial for a clear understanding of the meaning of the quantum state $| \psi \rangle$. Although the following brief review is quite standard, we shall nevertheless present it for the sake of drawing attention to our claim. \\

Limiting ourselves to non-relativistic quantum mechanics, we may consider spatial  $(\hat q, \hat p)$ and spin $\hat {\vec S}$ degrees of freedom only. For the case of spatial degrees of freedom, these obey the commutation rule, 
\begin{equation}
\left[\hat q, \hat p\right] = i \hbar .
\end{equation}
The operators $\hat q$ and $\hat p$ may represent position and momentum observables of a one dimensional spatial degree of freedom, with continuous eigenvalues $q \in \mathbb{R}$ and $p \in \mathbb{R}$, and with an associated Hilbert space ${\cal H}_{q,p}$. If the system is composed of this degree of freedom only, then a pure state $|\psi \rangle$ requires the specification of only one observable or operator, namely, of any Hermitian function $\hat f = f(\hat q,\hat p)$ with eigenvalues $f_n$. That is, any pure state is of the form $|\psi \rangle =  |f_n \rangle$, with $\hat f |f_n \rangle = f_n |f_n \rangle$ for some value of $n$. For a single spin degree of freedom, the commutation rule is 
\begin{equation}
\left[\hat S_j, \hat S_k \right] = i \hbar \epsilon_{jkl} \hat S_k ,
\end{equation}
where $(j,k,l)$ are cartesian components and $\epsilon_{jkl}$ is the fully antisymmetric Levi-Civita tensor. There is also an associated Hilbert space ${\cal H}_S$, but now the state requires the specification of two eigenvalues of two commuting observables, say of $\hat S^2$ and $\hat S_z$, and an example is $|\psi \rangle = |s,m \rangle$. \\

However, if the system has more than one degree of freedom, say one spatial and one spin degrees of freedom, a state may be $|\psi \rangle = |f_n, s, m\rangle$ indicating that the observables $\hat f$, $\hat S^2$ and $\hat S_z$ have the values $f_n$, $s$ and $m$ with certainty. The set of operators $\left\{\hat f, \hat S^2, \hat S_z\right\}$, that commute among themselves, is called a complete set of commuting observables (CSCO). In the joint Hilbert space ${\cal H}_{q,p} \otimes {\cal H}_S$ there does not exist any other operator that commutes with all the operators of the given CSCO, unless one trivially considers an operator that it is a function of them. On the other hand, there is an infinite number of different CSCO's in the joint Hilbert space. This result can be generalized to any number ${\cal N}$ of spatial and any number ${\cal M}$ of spin degrees of freedom. Any arbitrary state of the joint Hilbert space must be an specification of, at most \cite{aclara}, ${\cal N} + 2 {\cal M}$ eigenvalues yielding certainty on the values of the same number of operators conforming a CSCO. Of course, the states do not need to be product states, as the simple example above, but they can be entangled as sums of product states. The Hamiltonian of the system may or may not depend on all degrees of freedom. According to the present discussion, all states of a given system, at any time, are part of the complete set of states of some CSCO, indicating their certainty values; call such a state $|\psi \rangle = | \lambda_1, \lambda_2, \dots, \lambda_{{\cal N} + 2 {\cal M}} \rangle$. The values of any other CSCO, drawn from its complete set of states,  is undetermined, but one has a statistical knowledge of them provided by the transition probabilities $|\langle \zeta_1, \zeta_2, \dots, \zeta_{{\cal N} + 2 {\cal M}} | \lambda_1, \lambda_2, \dots, \lambda_{{\cal N} + 2 {\cal M}}\rangle|^2$, with $ | \zeta_1, \zeta_2, \dots, \zeta_{{\cal N} + 2 {\cal M}}\rangle$ a state of the other CSCO. To summarize, a  {\it state} of a {\it given} system amounts to the specification of one eigenvalue of each one of all the operators of a CSCO, and for which such an state yields full certainty. This is the meaning of the state $|\psi \rangle$ and its knowledge is the most complete specification of the state of a system given by quantum mechanics. Two questions remain, one is how we understand this completeness in the ``real" world and, second, how one can know in which state a system is.\\

To find in which state a system is, the recipe is to perform a ``measurement'' of the physical variables corresponding to the operators of any CSCO. Truly, this is not part of the theory, just as in classical mechanics there is no recipe as to how to produce the initial values of the variables of the given system. Certainly, in order to contrast the theory with the real world, some kind of measurement must be made, but the theory is silent as to how to perform it. That is actually our problem. Nevertheless, we must assume a measurement is performed such that one achieves knowledge with certainty that an observable takes on some of its eigenvalues. However, since full specification of a state requires the knowledge of, at most \cite{aclara}, ${\cal N} + 2 {\cal M}$ eigenvalues of a CSCO, one may be inclined to believe that ${\cal N} + 2 {\cal M}$ different measurements are necessarily needed. This is not true in general. It may be that we just need a number of measurements smaller than the number of operators in the CSCO. For purposes of the discussion, we define here a {\it complete measurement} as the measurement of the minimum number of observables needed to completely specify all the eigenvalues of a state of a CSCO. This minimum number cannot be universally specified, it depends on (a) the system itself; (b) the CSCO set we are interested in; and (c) the state we have certainty the system was at a previous time. There is no general procedure to establish how the knowledge of a state should be acquired. Before giving some examples, some of them of relevance to the issue discussed in this paper, we emphasise that the determination of the state of the system is considered complete when all the eigenvalues of a CSCO are known, because by hypothesis of the theory, namely, by the Uncertainty Principle, it must be impossible to construct devices that would simultaneously measure the needed observables of two or more distinct CSCO's. This, so far, has also hold true in real experimental situations.\\

Returning to the discussion of complete measurements, a simple illustrative example may be a spinless three-dimensional particle of mass $m$ in a central potential $\hat V = V(|\hat r|)$. The system has three spatial degrees of freedom. A well-known CSCO is given by $\{\hat H, \hat L^2, \hat L_z\}$, with $\hat H$ the Hamiltonian and $\hat {\vec L}$ the angular momentum vector. The states of this CSCO may be labeled as $|\psi \rangle = |n,l,m\rangle$ in known notation. We have, at least, three cases. First, if the potential is Coulombic and attractive, there is a full degeneracy that requires the measurement of the three operators in the CSCO to determine the state. Second, if the potential is arbitrary but not Coulombic, one may need $\hat H$ and $\hat L_z$ only, $\hat L^2$ is already determined and there is no need of measuring it. Third, if in addition to the potential being arbitrary there is a uniform arbitrary magnetic field in the $z$-direction, it may be necessary to measure $\hat H$ only, since the values of $\hat L^2$ and $\hat L_z$ are both determined by the measurement of $\hat H$. The point we want to make is that the completeness of the state requires the knowledge of the three eigenvalues $n$, $l$ and $m$, but the necessity of measuring all of them or or not, depends on the system and on the chosen CSCO. The latter condition is clear since any other arbitrary CSCO may need other type of measurements. For instance, consider the CSCO set of the position coordinates $\{\hat x,\hat y, \hat z\}$. In this case we always need to measure all of them, regardless of the potential and the presence or not of a magnetic field, in order to have a complete specification of the state. The other case of a complete measurement, and the most important for us here - that of the need of measuring a reduced number of observables because one knows the state the system is or was - will be discussed in the light of entangled states below. \\

{\it Entangled spin states.} Let $\{ |\pm, {\bf a} \rangle\}$ be the basis set of one spin in the projection $\hat S_a^{} = {\bf a} \cdot \hat {\vec S}^{}$, with ${\bf a}$ a given unit vector in 3-dimensional space. These states can written as,
\begin{equation}
|\pm , {\bf a} \rangle = e^{i \hat \sigma_z \phi/2} e^{-i \hat \sigma_y \theta/2} e^{- i \hat \sigma_z \phi/2} | \pm \rangle ,
\end{equation}
where ${\bf a}$ points along the direction $(\theta, \phi)$ in usual spherical coordinates and $|\pm \rangle$ is the basis set of $\sigma_z$. Consider now a system of two distinguishable spin degrees of freedom, each of spin 1/2, labeled as $\hat {\vec S}^{(1)}$ and $\hat {\vec S}^{(2)}$ with $s^{(1)} = s^{(2)} = 1/2$. Any CSCO of this system is given by 4 operators. However, since the total spin of each one is fixed, the CSCO's can be considered of only two operators. Any given state of the two-spin system can be written as
\begin{equation}
| \psi \rangle = \alpha |+,{\bf a} \rangle_1 |-; {\bf b} \rangle_2 -  \beta  |- ,{\bf a}\rangle_1 |+; {\bf c} \rangle_2 \label{enta2}
\end{equation}
with $|\alpha|^2 + |	\beta|^2 = 1$ and ${\bf a}$ an arbitrary direction. The directions ${\bf b}$ and ${\bf c}$ are determined by the state $|\psi \rangle$ and the chosen direction ${\bf a}$. The state is entangled if ${\bf b} \cdot {\bf c} \ne -1$ and both $\alpha \ne 0$ and $\beta \ne 0$.\\

The main characteristic of an entangled two-spin state is that the measurement of any of the two spins in any arbitrary direction ${\bf a}$ is a complete measurement. This is a very strong statement since it establishes that the measurement of the projection of the other spin is uniquely determined. More than that, the measurement of 
any other direction of the second spin does not commute with the previous one and, thus, {\it it cannot be done}. Let us use the state given by Eq.(\ref{enta2}) as an example and assume the two-spin state is such a $|\psi \rangle$. Assume we choose to measure spin 1 along ${\bf a}$, namely $\hat S_{a}^{(1)}$ and obtain $+1$. The main claim of this article is that this measurement is a {\it complete} measurement of the joint operator $\hat S_{a}^{(1)} \otimes \hat S_{b}^{(2)}$. If $-1$ is obtained instead, then, this is now a {\it complete} measurement of $\hat S_{a}^{(1)} \otimes \hat S_{c}^{(2)}$. Therefore, the pretension of measuring, say, $\hat S_{a}^{(1)} \otimes \hat S_{d}^{(2)}$ with ${\bf d } \ne {\bf b}$ and ${\bf d } \ne {\bf c}$, cannot be done. It violates the Uncertainty Principle. At first sight this sounds strange, or even incorrect, since it seems that {\it we} have the freedom to choose the directions ${\bf a}$ for spin 1 and ${\bf d} \ne {\bf b}$ or ${\bf d} \ne {\bf c}$ for spin 2, giving us the illusion that we are measuring the corresponding directions independently of each other. We argue below that the solution to this difficulty is that we can claim that, always, we do measure one of them first and later on the second one. That is, the measurement of two entangled spins at arbitrary directions is always decomposed into two successive complete measurements. If the state is not entangled there is no conflict, of course. The interesting consequence of this viewpoint is that, using the arguments put forward by Bell, the mentioned two complete measurements do not violate the criterion of locality. Put it in the terms we are advocating here, we find that the main characteristic of an entangled state is that {\it their} complete measurements and their associated CSCO's, depend on the measurement of only one of the pair and on the entangled state itself,  and not upon our apparent choice of both directions. Indeed, any product $\hat S_{a}^{(1)} \otimes \hat S_{d}^{(2)}$, for ${\bf a}$ and ${\bf b}$ arbitrary is a CSCO of all not-entangled states but of only certain entangled ones. As stated in the Introduction, this requires an assessment of the concept of simultaneity in quantum mechanics. We must conclude, however, that this quality cannot be precisely established since it requires of the precise elucidation of  the {\it velocity}, of a particle or of a reference frame, a classical concept in conflict with the Uncertainty Principle.  In many practical cases this delicate point may be ignored but not in this one.\\

{\it Bell theorem.} To briefly revise Bell criterion, let us limit ourselves to Bell state, originally proposed by Bohm \cite{Bohm}, 
\begin{equation}
|\psi_B \rangle = \frac{1}{\sqrt{2}} \left( |+ \rangle_1 |- \rangle_2 -   |- \rangle_1 |+ \rangle_2 \right). \label{BellS}
\end{equation}
Following Bell, we can consider the most general measurement of arbitrary projections of the two spins, namely, ${\bf a}\cdot \hat {\vec \sigma}^{(1)} \otimes {\bf b} \cdot \hat {\vec \sigma}^{(2)}$, with ${\bf a}$ and ${\bf b}$ arbitrary unit vectors, and $\hat {\vec S} = (\hbar/2) \hat {\vec \sigma}$. The expectation value of a very large number of similar measurements yields,
\begin{equation}
\langle \psi_B |\> {\bf a}\cdot \hat {\vec \sigma}^{(1)} \otimes {\bf b} \cdot \hat {\vec \sigma}^{(2)} \>|\psi_B \rangle = - {\bf a} \cdot {\bf b}. \label{PQ}
\end{equation}
Bell analyzed a hypothetical experimental situation where spin 1 and 2 ``fly" apart a very large distance, such that space-like measurements of both spins projections can be made. Without necessarily assuming that quantum mechanics holds, Bell considers the correlations of the obtained measured results $P({\bf a}, {\bf b})$, given by,
\begin{equation}
P({\bf a}, {\bf b}) = \int d\lambda \> \rho(\lambda) \>A({\bf a},\lambda) \>B({\bf b},\lambda) \label{PBell}
\end{equation}
where $A({\bf a},\lambda)$ and $B({\bf b},\lambda)$ are the results of the measurements of the projection of the spins 1 and 2, respectively. Both variables can only yield $\pm 1$ but they certainly depend on the orientations ${\bf a}$ and ${\bf b}$ and on an unknown number of additional or ``hidden" variables represented by the parameter $\lambda$, which is assumed to be given by a distribution $\rho(\lambda)$. The crucial observation of the correlation given by Eq.(\ref{PBell}) is that it obeys locality, in the sense of special relativity. That is, variable $A({\bf a},\lambda)$ does not depend on ${\bf b}$ and $B({\bf b},\lambda)$  does not depend on ${\bf a}$. The dependence en $\lambda$ can be restricted to depend on previous events, limited by the light-cones of the separated measurements, as Bell also showed \cite{Bell2}, making the proof more rigorous. The point is to enquire if it is possible to find a distribution function $\rho(\lambda)$, independent of the values of ${\bf a}$ and ${\bf b}$, such that $P({\bf a}, {\bf b}) = - {\bf a} \cdot {\bf b}$, or in other words, if the correlation given by Eq.(\ref{PBell}) can be made equal to the quantum expectation value given by Eq.(\ref{PQ}). The answer is {\it no} for arbitrary directions of ${\bf a}$ and ${\bf b}$. This result can be mathematically verified using any of the Bell inequalities \cite{Bell1,Ineq}. The impossibility of matching the quantum expectation value of a joint measurement with the corresponding correlation function, imposing locality for arbitrary directions ${\bf a}$ and ${\bf b}$, is Bell theorem. As it has been extensively claimed, experiments not only violate Bell inequalities, their results match the quantum prediction. In reviewing the origin of the discrepancies, it has been concluded that it is the assumed {\it spatial locality} what it is not obeyed in space-like separated measurements of the given entangled state. We now argue that this conclusion does not necessarily follow.  \\

{\it The quantum loophole.} Our contention is based on arguing that the comparison between the correlation function Eq.(\ref{PBell}) and the quantum expectation value Eq.(\ref{PQ}) is not valid. Let us first use a result of Bell \cite{Bell1} that, although mentioned in many discussions of Bell inequalities, we believe it has not been thoroughly exploited. Bell showed that if the measured projections are parallel, namely ${\bf a} = {\bf b}$ (and also for ${\bf a} \cdot {\bf b} = 0, -1$), one can consider a uniform distribution for $\rho(\lambda)$ that yields the corresponding quantum expectation value. Not only that, for the measurement of a single spin, Bell also showed that a local ad-hoc hidden variable distribution can always be found. Although we have already spelled out the alternative interpretation of the pair measurements, we readdress it below.\\

Regard a situation where, on purpose, the measuring devices are arranged to perform space-like measurements of  the projection of spin 1 at a time $t_1$ and  of the projection of spin 2 at a time $t_2 = t_1 + \tau$, with $\tau$ a finite time delay. Then, following our discussion above, we claim that the measurement of spin 1 constitutes a complete measurement of the pair. That is, if spin 1 yielded $-1$ at orientation ${\bf a}$, then, we know with certainty that spin 2 has the value $+1$ at the same orientation ${\bf a}$. We reiterate the strong statement that for an orientation ${\bf a}$ of spin 1, we can {\it only} measure ${\bf a}$ of spin 2 without interfering with the Uncertainty Principle, and therefore, spin 2 does not need to be measured. If we did it along ${\bf a}$, we would certainly obtain the same result. In this case, and only in this one, we can compare Bell correlation $P({\bf a},{\bf a}) = -1$ with the expectation value of Eq.(\ref{PQ}) of many similar experiments and we do find agreement. As stated above, this case does not violate locality. \\

Now, consider the case where the projection of spin 2 is measured at orientation ${\bf b} \ne {\bf a}$ at time $t_2 = t_1 + \tau$, having obtained, say a value $-1$ for the projection at ${\bf a}$ of spin 1 at time $t_1$. Hence, we can assert that spin 2 has the value $+1$ at orientation ${\bf a}$ at time $t_1$ and, therefore, it faces the measurement at time $t_2 = t_1+\tau$ at ${\bf b}$ being already {\it prepared} in such a state. Its result will be $+1$ or $-1$ weighted by the corresponding transition probabilities ( $|_2\langle {\bf a}, +1| {\bf b}, +1\rangle_2|^2$ or $|_2\langle {\bf a}, +1| {\bf b}, -1\rangle_2|^2$) but, as Bell showed, this neither violates locality. The two measurements of spin 2 are certainly time-like separated. Many repetitions of this composite measurement yield the same expectation value as that given by Eq.(\ref{PQ}). However, we can no longer compare with Bell correlation function for they refer to different situations. Bell correlation assumes at best a common past when the entangled state was created and that {\it nothing} interfered with the previous courses to the separated experimentes, except a set of unknown or hidden variables. The view point of quantum mechanics indicates that the measurement of spin 1, being a complete measurement, terminates the evolution initiated with the entangled state. This measurement serves to yield an ``initial" state of spin 2 that it is further measured along ${\bf b}$. In other words, the measurement of spin 2 at time $t_2 = t_1 + \tau$ along ${\bf b}$ is no longer related, or ``correlated", with the original entangled state.\\

The loophole can now be simply stated. The key is that the delay time  $\tau$ is finite, although it could be made as small as we wish. We claim that in real experiments with pairs of photons instead of spins\cite{AspectEPR,expEPR}, it cannot be assured that the measurements at detectors ${\bf a} \ne {\bf b}$ are actually simultaneous. These occur within very small, yet finite time windows and, in some cases, such as in two-photon cascades \cite{AspectEPR}, the photons are certainly not emitted simultaneously but at undetermined different times. In parametric down conversion emission, while the pair emission could be considered to be simultaneous, there is a quantum uncertainty regarding the precise spatial location of the emission, and we can further argue, of the detection. Therefore, we assert, even within short coincidence time-window detections, one photon is detected first, the entanglement is finished yielding certainty on the value of the other - without violating locality - then the second faces later another detector but in an already known state, just as described above. \\

{\it A comment on the measurement process.} Although there is no pretense here in addressing how an actual measurement occurs, it is certainly clear that what it is at stake here is the understanding of the measurement process in quantum mechanics. Referring to the case at hand, by admitting that the entangled pair measurement is achieved by measuring only one of them {\it and} by knowing that the state is entangled, we can no longer assert that it is necessarily the presence of a measuring device what produces certainty on the value of an observable. We also exemplified this situation with the 3-dimensional particle in different external potentials. Thus, we can say that, although the presence of some measuring device is completely necessary, the concept of the ``collapse" of the state at contact with the measuring device appears dubious or simply unnecessary. This view point by no means is new. Although not stated explicitly in his reply to EPR \cite{EPR}, Bohr \cite{Bohr} does not embrace the concept of the collapse, on the contrary, he insists that the measuring device does not affect the measured variable but rather its presence makes it impossible to determine the value of the variables not measured. This position of Bohr has also been pointed out by some phylosophers of science {\cite{Faye}}. In the opinion of the author, the necessity of the concept of the collapse arises because of the interpretation of the superposition of the states as something tangible or measurable. However, as we have expressed here, the meaning of the state $|\psi \rangle$ is its certainty on eigenvalues of the CSCO it represents.  Its elucidation needs a complete measurement which does not necessarily requires the presence of devices for the values of all variables to be found. Therefore, if one considers the superposition of states as a mere mathematical device to predict probable outcomes of an experiment, as many authors have also insisted, the need of the collapse becomes also superfluous. Again, while not trying to ``explain" how the process occurs, one may say that the purpose of a measurement is to unveil the value of an observable, and the least it affect it, the better. The Uncertainty Principle takes care of the fact that we can find eigenvalues of a single CSCO at a given time only.\\

{\it Acknowledgement.} The author thanks the attendees of the seminar {\it Fundamenta Quantorum}, at the Institute of Physics, UNAM, for the fruitful discussions on subjects related to this article. The attendees of the seminar do not necessarily share the viewpoints here exposed.


\begin{thebibliography}{99}
\bibitem{aclara} Considering a system with ${\cal N}$ spatial degrees of freedom only, the number of commuting operators in any CSCO cannot be larger than ${\cal N}$. However, it may be smaller. While full elucidation of this point is beyond the scope of this paper, one may conclude such a result by appealing to the fact that a  partially non-integrable classical Hamiltonian may have a  number of integrals of the motion in involution smaller than ${\cal N}$, being equal to ${\cal N}$ if integrable \cite{Perez}. Thus, translating this result into the quantum case one may then assert that there exist CSCO's with a number of operators smaller than ${\cal N}$. This does not affect the definition of ``complete measurement", since the latter needs a number of measurements equal or smaller than  the number of operators in the corresponding CSCO.
\end{thebibliography}
\end{document}